\documentclass[%
 aip,
 amsmath,amssymb,
 reprint,%
]{revtex4-1}
\usepackage{xcolor}
\usepackage{graphicx}
\usepackage{dcolumn}
\usepackage{bm}

\usepackage[utf8]{inputenc}
\usepackage[T1]{fontenc}
\usepackage{mathptmx}
\usepackage{subcaption}
\usepackage{ifthen}
\begin{document}

\preprint{AIP/123-QED}

\title{Resolving the structure-energy dilemma at organic-inorganic interfaces: Adsorption of benzene, thiophene and xenon over coinage metal surfaces}

\author{Santosh Adhikari}
\email{tuf60388@temple.edu.}
\author{Niraj K. Nepal}
\author{Hong Tang}
 \affiliation{Department of Physics, Temple University, Philadelphia, PA-19122, USA}
 
\author{Adrienn Ruzsinszky}
\affiliation{Department of Physics, Temple University, Philadelphia, PA-19122, USA}

\date{\today}

\begin{abstract}
    Semilocal (SL) density functional approximations (DFAs) are widely applied but have limitations due to their inability to incorporate long-range van der Waals (vdW) interaction. Non-local functionals (vdW-DF, VV10, rVV10) or empirical methods (DFT+D, DFT+vdW, DFT+MBD) are used with SL-DFAs to account for such missing interaction. The physisorption of a molecule on the surface of the coinage metals (Cu, Ag, and Au) is a typical example of systems where vdW interaction is significant. However, it is difficult to find a general method that reasonably describes both adsorption energy and geometry of even the simple prototypes of cyclic and heterocyclic aromatic molecules like benzene (C$_6$H$_6$) and thiophene (C$_4$H$_4$S) respectively, with reasonable accuracy. In this work, we present an alternative scheme based on Zaremba-Kohn theory, called DFT+vdW-dZK. We show that, unlike other popular methods, DFT+vdW-dZK and particularly SCAN+vdW-dZK gives an accurate description of the physisorption of a rare-gas atom (Xe) and two small albeit diverse prototype organic molecules on the (111) surfaces of the coinage metals. 
\end{abstract}

\maketitle

\section{Introduction}
The interface between an organic molecule and a metallic surface plays a significant role in diverse fields such as optoelectronics, catalysis, sensors, and surface-photochemistry.\cite{koch2013molecule,oehzelt2014organic,witte2004growth,friend1997fundamental,gomez2001hybrid} Benzene (C$_6$H$_6$) and thiophene (C$_4$H$_4$S) are respectively the most studied\cite{takahashi2002new,revelli2010extraction,adhikari2020fermi} prototypes of the cyclic and the heterocyclic aromatic molecules to model\cite{reckien2014theoretical,christian2016surface,tkatchenko2009accurate,tkatchenko2012accurate} such an interface. There are experimental techniques like temperature-programmed desorption (TPD), normal incidence X-ray standing wavefield absorption (NIXSW), and near-edge X-ray absorption fine structure spectroscopy (NEXAFS) to help determine the adsorption energy and geometry at the interface. However, experimental results are affected by surface defects, monolayer formation, the orientation of the adsorbate, etc., and require analysis from several experiments to be put together for the full picture.\cite{witte2004growth} These difficulties motivate the need for computational tools that are able to mimic the mechanism that requires a delicate balance between the short- and long-range interactions (mainly van der Waals (vdW) effect).\cite{tkatchenko2010van,liu2014modeling,stohr2019theory,yuan2020effects} Being a nonlocal effect, the latter poses a tough challenge for theoretical methods.\\
$~~~~$Wave-function based approximations like coupled-cluster (CC) can account for the vdW effect with greater accuracy but are computationally out of reach for such interface models. Alternatively, density functional theory (DFT)\cite{hohenberg1964inhomogeneous,kohn1965self}  provides a framework for electronic structure calculations in diverse fields\cite{jones2015density} through different rungs of approximations illustrated by the Jacobs ladder of Perdew.\cite{jacobsladder2001} High-level approximations like the random phase approximations (RPA) within this framework are nearly exact for the long-range vdW effect.\cite{dobson2005soft} But, despite the increased computational power and efficient implementation,\cite{nguyen2010first} RPA is unable to handle supercell calculations involving a relatively large numbers of atoms. DFT can accomodate these calculations at the semilocal (SL) level, but the required long-range correlation effects are missing here. Therefore, SL density functional approximations (DFAs) are often combined with a nonlocal correlation functional\cite{dion2004van,vdwdf2,vydrov2010nonlocal,rvv10} or an empirical method\cite{becke2005density,johnson2005post,becke2007unified,grimme2004accurate,grimme2006semiempirical,grimme2010consistent,grimme2011effect,tkatchenko2009accurate,ruiz2012density,tkatchenko2012accurate} that relies on input polarizabilities, mostly through some adjustable parameters to include the vdW effects. These adjustable parameters guide the delicate balance between the short-range interactions captured by DFAs (base functionals) and the long-range interactions that the correction methods provide.\cite{klimevs2012perspective}\\
$~~~~$GGAs (generalized gradient approximations) are superior to LDA for aromatic systems\cite{jenkins2009aromatic} and hence are the natural choices as base functionals. PBE\cite{pbe1996} and its revised versions\cite{revPBE,perdew2008restoring,optPBE} and the Becke functionals\cite{B86,B88,B86b} are some successful\cite{yildirim2013trends,christian2016surface,reckien2014theoretical,maass2018binding,liu2013structure,peng2017rehabilitation,terentjev2018dispersion,adhikari2020molecule} base functionals at the GGA level. Despite being widely used, GGAs by themselves are unable to capture vdW interaction beyond the short-range while the SCAN\cite{scan2015} metaGGA does. SCAN, by construction, can recognize different bonding environments. There are a few instances like formation energies of weakly bound intermetallics,\cite{isaacs2018performance,nepal2019treating,nepal2020formation} and prediction of magnetic properties of transition metals,\cite{fu2018applicability} where SCAN is not up to the mark, but it has been mostly successful for diversely bonded systems.\cite{sun2016accurate,shahi2018accurate,nepal2018rocksalt,nepal2019first,yu2020different} Recently, SCAN was revised (revSCAN\cite{mezei2018simple}) by diminishing its intermediate-ranged vdW interaction, leading to better interaction energies of molecules of the S22 dataset when combined with VV10\cite{vydrov2010nonlocal} than SCAN+VV10. However, when combined with rVV10,\cite{rvv10} SCAN is remarkably accurate for molecular complexes, solids, and layered materials.\cite{peng2016versatile}\\
Both VV10 and rVV10 are nonlocal correlation functionals like vdW-DFs,\cite{dion2004van,vdwdf2} paired with the base functional as
\begin{equation}
    E_{xc}=E{_{xc}}{^{DFA}}+E{_{c}}{^{nl}},
\end{equation}
where E${_{xc}}{^{DFA}}$ is the total exchange-correlation energy computed from the base functional, while E${_{c}}{^{nl}}$ is the nonlocal electron-correlation energy. These schemes bridge the nonlocal correlation functionals to the semilocal functional through a few parameters and depend only on the electron density as input.\\
$~~~~$There is an alternative way to include the vdW interaction through empirical methods where the base functional is often combined  as
\begin{equation}
    E_{tot} = E_{DFA} + E_{vdW},
\end{equation}
where E$_{DFA}$ is the total energy computed from the base functional while E$_{vdW}$ is the contribution to the total energy due to vdW interaction. Methods like DFT+D\cite{dft+d,dft+d2} model the latter part as a C$_6$ coefficient-based pairwise additive interaction. The C$_6$ coefficients utilized here are unable to adjust to the chemical environment (system) and neglect the many-body dispersion effects. But methods like DFT+D3,\cite{dft+d3-zero,dft+d3-bj} XDM\cite{becke2005density,johnson2005post,becke2007unified}(exchange-hole dipole moment), and DFT+vdW\cite{tkatchenko2009accurate} incorporate the information of the system through different ways.\cite{klimevs2012perspective}\\
$~~~~$In our present assessment, we are investigating the adsorption of benzene, thiophene, and xenon over the (111) surfaces of the coinage metals. Several past studies have suggested the physisorption of these systems on such coinage metal surfaces. However, it is difficult for even the improved empirical functionals to model such an interface due to the screening of molecule-surface interaction.\\
$~~~~$We present in the Methodology section a brief introduction of the recently proposed model\cite{tao2018modeling} based on Zaremba-Kohn's second-order perturbation theory designed for modeling the physisorption of a particle over the surface that naturally incorporates the screening effects. This model yields RPA quality results for the physisorption of graphene over metallic surfaces\cite{tao2018modeling,tang2018} and semiconducting layered materials. In our previous work,\cite{adhikari2020molecule} we showed that this method outperforms all other semilocal functionals paired with rVV10 for adsorption of thiophene on coinage metal surfaces. We note that the model DFT+vdW\cite{ruiz2012density} is based on Lifshitz-Zaremba-Kohn's theory, too. However, that model includes the screening effects through the C$_6$ instead of C$_3$ coefficients and has no high order terms, such as quadrupolar C$_5$.\\
\section{Methodology}
A recently introduced model\cite{tao2018modeling} based on the damped Zaremba-Kohn\cite{lifshitz1956theory,zaremba1976van,zaremba1977theory} (dZK) second-order perturbation theory addresses the vdW interaction for physisorption of graphene over a metallic surface. In this model, the asymptotic form of the vdW interaction between a particle and a clean surface accounts\cite{jiang1984dispersionlocal,jiang1984dispersionnonlocal} for vdW energy as
\begin{equation}
E_{vdW} = [-\frac{C_3}{(z-z_{0})^3}-\frac{C_5}{(z-z_{0})^5}]f_d.
\end{equation}
In Eq. (3), $z_0$ is the reference plane position and $z = d - c/2$, where $d$ is the distance between the particle and the surface and $c = a /\sqrt{h^2 + k^2 + l^2}$ with $h$, $k$, and $l$ being the Miller indices of the surface of transition metal having lattice constant $a$. Here the $C_3$ coefficient describes the dielectric response of the bulk solids to the instantaneous dipole moments of the particle, and the $C_5$ coefficient gives the fluctuating quadrupole (C$_{5}^{q}$), nonlocal (C$_{5}^{nl}$) and diffuse (C$_{5}^{d}$) contributions of the particle (see Tao \textit{et al}.\cite{tao2018modeling}  and references therein for more details). In the dZK model, the dynamic dielectric response from the substrates includes the screening effects. The C$_3$ and C$_{5}^{q}$ terms here are portrayed as a kind of local form, in which the substrate response is approximated in terms of an average dielectric function, computed at wave vector q = 0. The C$_{5}^{d}$ and C$_{5}^{nl}$ terms are derived from the hydrodynamic model at electrostatic limits, and are dipolar contributions to the term $\sim$ 1/${(z-z_{0})^5}$. The wave vector ( q$\neq$0 ) dependence is included in the derivations of C$_{5}^{d}$ and C$_{5}^{nl}$.\cite{jiang1984dispersionlocal,jiang1984dispersionnonlocal} Those terms include effects of a nonlocal dielectric function. This model relies on the highly accurate static polarizabilities from experiment\cite{tandfonline,nist} or from high-level calculations.\cite{smith2004static,delaere2002influence,withanage2019self}
The damping factor for Eq. (3) is
\begin{equation}
f_d=\frac{x^5}{(1+gx^{2}+hx^{4}+x^{10})^{\frac{1}{2}}}  
\end{equation}
where, $x=\frac{z-z_{0}}{b} > 0$, $g=\frac{2 b^{2}C_3}{C_5}$ and $h=\frac{10 b^{4}{C_3}^{2}}{{C_5}^{2}}$. The cutoff parameter $b$, which is our only fit parameter, is determined by minimizing the mean absolute error (MAE) between calculated and RPA binding energies of graphene over the transition metal surfaces.\cite{tao2018modeling,tang2018} Based on previous work,\cite{tang2018,adhikari2020molecule} we used b=3.3 Bohr for PBE+vdW-dZK. Here we are using b=4.1 Bohr for SCAN+vdW-dZK.
Instead of taking the static dipole polarizability of the molecule, we base our $C_3$ coefficients on the “renormalized atom” approach.\cite{tang2018} The best polarizability for a particular atom (H, C, or S) in thiophene or (H or C) in benzene is then renormalized as   
\begin{equation}
\alpha_{(renormalized\_atom)}=\frac{\alpha_{(free\_atom)}}{4{\alpha_{(C)}}+4{\alpha_{(H)}}+{\alpha_{(S)}}} \alpha_{(thiophene)}
\end{equation}
for thiophene, and
\begin{equation}
\alpha_{(renormalized\_atom)}=\frac{\alpha_{(free\_atom)}}{6{\alpha_{(C)}}+6{\alpha_{(H)}}} \alpha_{(benzene)}
\end{equation}
for benzene.\\
$~~~~$With the static polarizabilities, we can find the separate $C_{3}$ and $C_{5}$ coefficients for each of the elements in the molecule (see Tables (S1-S3) in Supplementary Materials for the computed C$_3$ and C$_5$ coefficients for xenon, and renormalized atoms in benzene and thiophene along with the reference plane positions (Z$_0$)). The formula of renormalization we are using is for a "particle" interacting with a metal surface. Despite that both benzene and thiophene are relatively small molecules, we can not treat them as particles. Treating them as particles would overestimate $C_5$ significantly.\cite{adhikari2020molecule} Thus we treat benzene and thiophene as a collection of renormalized atoms. 
\section{Computational details}
All the DFT calculations performed utilized the projector augmented wave (PAW) formalism implemented in the Vienna ab initio simulation package (VASP) code. Geometry relaxations of the bulk structures of silver, gold, and copper using different XC functionals yielded the respective lattice constants (see Table S4 in Supplementary Materials for calculated lattice constants compared to the experimental zero-point phonon corrected lattice constants.\cite{hao2012lattice}) Since a study by Carter \textit{et al}.\cite{carter2014van} showed a non-negligible interaction of the adsorbate-molecule with its periodic image for a relatively smaller (3 $\times$ 3) supercell, we opted to build a (4 $\times$ 4) supercell of  111 surfaces in the atomic simulation environment (ASE)\cite{ase-paper,ISI:000175131400009} using the optimized lattice constants. The supercell has a five-atomic-layer thickness. A vacuum of 12 {\AA} was added along the z-direction to prevent the interactions due to the periodic images. The positions of the atoms on the bottom three layers were fixed during the relaxation to reduce the computational cost. Benzene and thiophene constructed using the reference C-S, C-C, and C-H bond lengths\cite{harshbarger1970electron,mootz1981crystal} were allowed to relax in the slab whose size was identical to that of the surface on which the adsorption occurred. Initially, both thiophene and benzene were placed in a parallel orientation 3{\AA} above the top metal layer and were allowed to relax on the high-symmetry sites (fcc, hcp, ontop, and bridge).\cite{ase-paper,ISI:000175131400009} Those sites were defined using the center of mass and the azimuthal angle of the molecule  as proposed by Liu \textit{et al}.\cite{liu2013structure} For example, hcp-45 indicates the center of mass of thiophene adsorbed at hcp site with a symmetry axis rotated by $45^0$ from the direction of metal rows. The surface, the molecule, and the molecule-surface system were all separately relaxed.
We utilized a similar procedure for the adsorption of xenon (Xe) over the (111) surfaces of Ag and Cu, as well, except that Xe being an atom was just placed above the aforementioned high-symmetry sites without rotation. We used the VASP recommended PAW pseudopotentials for all the calculations. While we utilized a plane-wave cutoff of 650 eV, the smearing parameter ($k_{B}T$) of 0.1 eV was used following the first order Methfessel-Paxton scheme. We used 4 $\times$ 4 $\times$ 1 and 20 $\times$ 20 $\times$ 20 Monkhorst-Pack meshes for the Brillouin zone sampling of the surface and the bulk, respectively. Since both adsorption energies and equilibrium distances depend on the adsorption-site, we utilized all major high symmetry sites for PBE+vdW-dZK and SCAN+vdW-dZK calculations. However, we only used the most stable site (hcp-30 for benzene and fcc-45 for thiophene) for the computation from other methods. We calculated the adsorption energy by subtracting the energy of the surface and molecule from the surface-molecule system.
\begin{equation}
E_{ad}= E_{surf + mol} - E_{surf} - E_{mol}
\end{equation}
We adopted similar procedure for calculating the adsorption energy of Xe over the metallic surfaces.\\
$~~~~$For DFT+vdW-dZK calculations, we computed the C$_3$, C$_{5}^{q}$, C$_{5}^{nl}$, and $z_0$ terms using Eqs (2-4) from the work of Tao \textit{et al}.\cite{tao2018modeling} Since the diffuse part (C$_{5}^{d}$) is small compared to the nonlocal and quadrupole parts,\cite{jiang1984dispersionnonlocal,tao2018modeling} we compute C$_5$ as a sum of the C$_{5}^{nl}$ and C$_{5}^{q}$ terms (see Tables (S1-S3) in Supplementary Materials for the computed C$_3$ and C$_5$ coefficients for xenon, and renormalized atoms in benzene and thiophene along with the reference plane positions ($z_0$)). All the parameters required for the transition metals were taken from the previous work.\cite{tao2018modeling,tang2018} We utilized the experimental static dipole polarizabilities reported in the NIST\cite{nist} database for all atoms and molecules studied in our work. Finally, we computed the total adsorption energy at each distance by adding the adsorption energy from DFT calculations and the corresponding correction obtained from vdW-dZK model (Eq (3)), following the binding-curve approach as in Fig. 1 of Tang \textit{et al}.\cite{tang2018} (see Figures (S1-S8) in Supplementary Materials for the computed adsorption energies for benzene, thiophene and Xe over the (111) surfaces of coinage metals using PBE+vdW-dZK and SCAN+vdW-dZK). 
\section{Results and Discussion}
We have assessed the adsorption energies, the vertical adsorption distances, and the tilt angles of benzene and thiophene placed over the (111) surfaces of Cu, Ag, and Au (coinage metals) using the scheme based on the physisorption model (vdW-dZK)\cite{tao2018modeling} following Zaremba-Kohn theory. Currently, this scheme\cite{tao2018modeling} has been implemented to PBE\cite{pbe1996} (PBE+vdW-dZK\cite{tang2018,adhikari2020molecule}) and very recently the modified\cite{tang2020density} form of that scheme to SCAN metaGGA. In this work we are implementing the original scheme\cite{tao2018modeling} (without any modifications), to SCAN\cite{scan2015} (SCAN+vdW-dZK). For comparative purposes, we also paired the semi-local functionals with the non-local functional (rVV10\cite{rvv10}) or the D3\cite{dft+d3-bj} method. The common goal of all these methods is to capture the van der Waals (vdW) interaction that is very relevant for the adsorption process but is missing in the bare semi-local functionals. We utilized the parameters of Peng \textit{et al}.\cite{peng2016versatile} based on fitting to binding energy curve of argon-dimer obtained from CCSD(T) calculations, to pair rVV10 with SCAN. In our previous work,\cite{adhikari2020molecule} we followed the same approach to determine the parameters required to combine rVV10 with PBE, PBEsol, and revSCAN. Finally, the parameters obtained from Brandenburg \textit{et al}.\cite{brandenburg2016benchmark} were utilized to combine SCAN with D3 (SCAN+D3).\\
\subsection{Benzene over Cu(111), Ag(111) and Au(111)}
We utilized eight different high symmetry sites for the adsorption of benzene over the (111) surfaces of Cu, Ag, and Au, as displayed in Fig 1.
\begin{figure}[h!]
    \centering
    \includegraphics[scale=0.3]{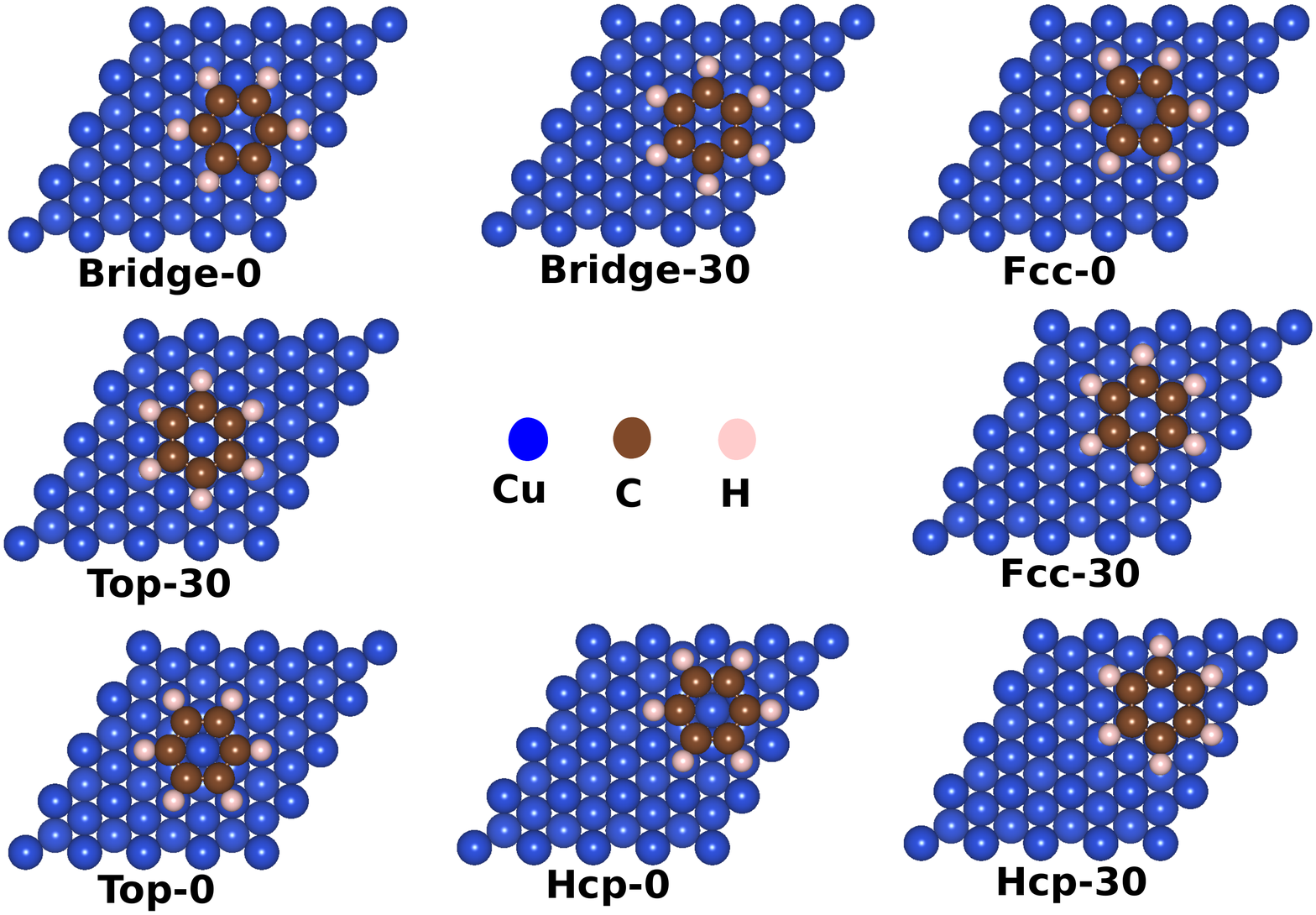}
    \caption{The figure shows the high symmetry sites\cite{ase-paper} utilized for the initial adsorption of a benzene molecule over the (111) surface of Copper.}
    \label{fig:my_label}
\end{figure}
Our calculations with PBE+vdW-dZK and SCAN+vdW-dZK, in agreement with Liu \textit{et al}.,\cite{liu2013structure} show a weak dependence of adsorption energy on the site as the maximum energy difference between them is just 0.05 eV. In agreement with past studies,\cite{bilic2006adsorption,peng2016versatile,liu2013structure,peng2017rehabilitation} these methods, irrespective of the substrate metals, predict the hollow site (hcp-30) as the most stable site of adsorption. For better comparison, we focused on the most stable site (hcp-30) for all other methods utilized in our work.\\
$~~~~$In agreement with the experiments,\cite{xi1994benzene,zhou1990interactions,syomin2001identification} all the methods discussed here including PBE+vdW-dZK and SCAN+vdW-dZK predict benzene to adsorb flat (parallel) to the metal surface without any noticeable tilting.\\
$~~~~$Table I displays the adsorption energy of benzene over the (111) surface of different metal substrates at the most stable site of adsorption. The experimental adsorption energies reported are acquired from the analysis of the data obtained from the temperature-programmed desorption (TPD) measurements. The most commonly used technique is the Redhead analysis,\cite{redhead1962thermal} which treats desorption temperature to be independent of the coverage and uses a frequency parameter often guessed over a range of values ([10$^{12}$ - 10$^{15}]$ s$^{-1}$). However, the TPD measurements display a correlation between the coverage and the desorption temperature, questioning the Redhead's analysis. A recent experimental study\cite{maass2018binding} utilizes the so-called complete analysis,\cite{king1975thermal} where the frequency factors are determined uniquely for a given coverage. This experiment\cite{maass2018binding} benchmarks the adsorption energy for benzene adsorbed (within an error bar of 0.05 eV) on the (111) surface of the coinage metals, enabling one to test the accuracy of theoretical methods.\\
\begin{table}[h!]
\caption{Adsorption energy (in eV) of benzene (C$_6$H$_6$) when placed on the most stable site (hcp-30) over the (111) surfaces of coinage metals using the different methods. The results from various previous studies are also reported here for comparison.}
\begin{tabular}{lcccc}
\hline \hline
              & C$_6$H$_6$/Cu(111) & C$_6$H$_6$/Ag(111) & C$_6$H$_6$/Au(111) &  \\
\hline
PBE           & -0.06      & -0.06      & -0.06      &  \\
PBE+rVV10     & -0.58      & -0.54      & -0.60      &  \\
PBE+D3        & -1.00      & -0.86      & -0.93      &  \\
PBE+vdW-dZK   & -0.56      & -0.60      & -0.67      &  \\
PBE+vdW\cite{carrasco2014insight}        & -1.07      & -0.87      & -0.84      &  \\
PBE+vdW$^{surf}$\cite{jiang2016aromatic}        & -0.86      & -0.75      & -0.74      &  \\
PBE+MBD\cite{maass2018binding}        & -0.63      & -0.57      & -0.56      &  \\
PBE+XDM\cite{christian2016surface}        & -0.54      & -0.58      & -0.61      &  \\
B86bPBE+XDM\cite{christian2016surface}        & -0.59      & -0.68      & -0.64      &  \\
optPBE+vdw-DF\cite{yildirim2013trends}        & -0.68      & -0.71      & -0.71      &  \\
SCAN          & -0.43      & -0.40      & -0.42      &  \\
SCAN+rVV10    & -0.78      & -0.75      & -0.80      &  \\
SCAN+D3       & -0.93      & -0.81      & -0.86      &  \\
SCAN+vdW-dZK  & -0.63      & -0.60      & -0.65      &  \\
revSCAN+rVV10 & -1.11      & -0.93      & -1.03      &  \\
HSE+MBD\cite{maass2018binding}       & -0.78      & -0.68      & -0.67      &  \\
Expt\cite{maass2018binding}           &  -0.68$\pm$0.04          &-0.63$\pm$0.05            &-0.71$\pm$0.03            & \\
\hline \hline
\end{tabular}
\end{table}
$~~~~$PBE largely underestimates the adsorption energies due to its inability to capture the vdW interaction beyond short-range. SCAN by design can incorporate some intermediate-range interaction, but as evident in our calculations, these are still not sufficient to fully account for the missing long-range vdW interaction. Non local correlation functionals like rVV10,\cite{rvv10} VV10\cite{vydrov2010nonlocal} and vdW-DFs\cite{dion2004van,vdwdf2} can often be paired with semilocal (SL) density functional approximations (DFA) to incorporate such missing interaction. However, our work suggests that the rVV10-based methods are inconsistent due to their over-flexibility in pairing with the base functionals. We will elaborate more on this puzzle in the later sections. There are alternative empirical ways to include vdW interactions such as the popular method DFT+D3,\cite{dft+d3-zero,dft+d3-bj} known for identifying the local chemical environment through improved C$_6$ coefficients. These methods, in comparison to their predecessors,\cite{dft+d,dft+d2} are much accurate for interaction energies of molecular systems like the S22 data set, rare-gas dimers, etc. But in agreement with the previous work,\cite{reckien2014theoretical} we found overestimated adsorption energies with these methods. Since the effect of surface metal atoms screening the molecule-surface interaction is ignored even in these improved C$_6$ coefficients, it could be the major reason behind the poor performance of these methods for our systems. In the DFT+vdW\cite{tkatchenko2009accurate} method, the C$_6$ coefficients along with the vdW radii are determined using the electron density obtained from the base functionals, and are independent of these underlying base functionals. However, like DFT+D,  this method also lacks screening effects leading to significantly overestimated\cite{carrasco2014insight} adsorption energies. The method DFT+vdW$^{surf}$\cite{ruiz2012density} was introduced to include the screening effects in DFT+vdW following Lifshitz-Zaremba-Kohn\cite{lifshitz1956theory,zaremba1976van} theory. Here, the screening effects were included using modified C$_6$ coefficients rather than explicitly using the C$_3$ coefficients pertinent for particle-slab interaction. PBE+vdW$^{surf}$ predicted better\cite{jiang2016aromatic} adsorption energies compared to PBE+vdW\cite{carrasco2014insight} but still the overestimation was noticeable. Another approximation, the DFT+MBD,\cite{tkatchenko2012accurate} was introduced to incorporate the screening effects in DFT+vdW. This method reported better adsorption energies\cite{maass2018binding} when combined with a hybrid functional (HSE). However, using a hybrid base functional is computationally demanding. Similarly the method PBE+XDM,\cite{christian2016surface} based on the exchange-hole dipole moment, \cite{becke2005density,becke2007unified} reports slightly underestimated adsorption energies. 
In Table I, we show the results from other important methods from previous studies,\cite{carrasco2014insight,christian2016surface,jiang2016aromatic,maass2018binding} as well.\\
$~~~~$Among all reported methods, SCAN+vdW-dZK stands out. The adsorption energies of benzene over different metal substrates predicted from this method lie close to the error bars of the reported reference values. In Fig. 2 we show the absolute deviation of the adsorption energies calculated using different methods in our work from the reported experimental values. Despite slightly underbinding benzene on a copper surface, PBE+vdW-dZK lies closest to the reference values for the other two metal surfaces (see Figures (S1-S3) in the Supplementary Materials for the details on PBE+vdW-dZK and SCAN+vdW-dZK calculations). The adsorption energies predicted from both vdW-dZK based methods are close to each other as was also the case for graphene over layered materials,\cite{tang2020density} which shows the consistency of these methods for adsorption energies. Recently, Chowdhury \textit{et al}.\cite{tanvir2021} have extended the vdW-dZK model to address the adsorption of molecules on carbon nanotubes, which have curved cylindrical surfaces. The binding energies calculated from PBE+vdW-dZK and SCAN+vdW-dZK are very close to each other. It further confirms the consistency of the vdW-dZK model.\\ 
\begin{figure}[h!]
    \centering
    \includegraphics[scale=0.6]{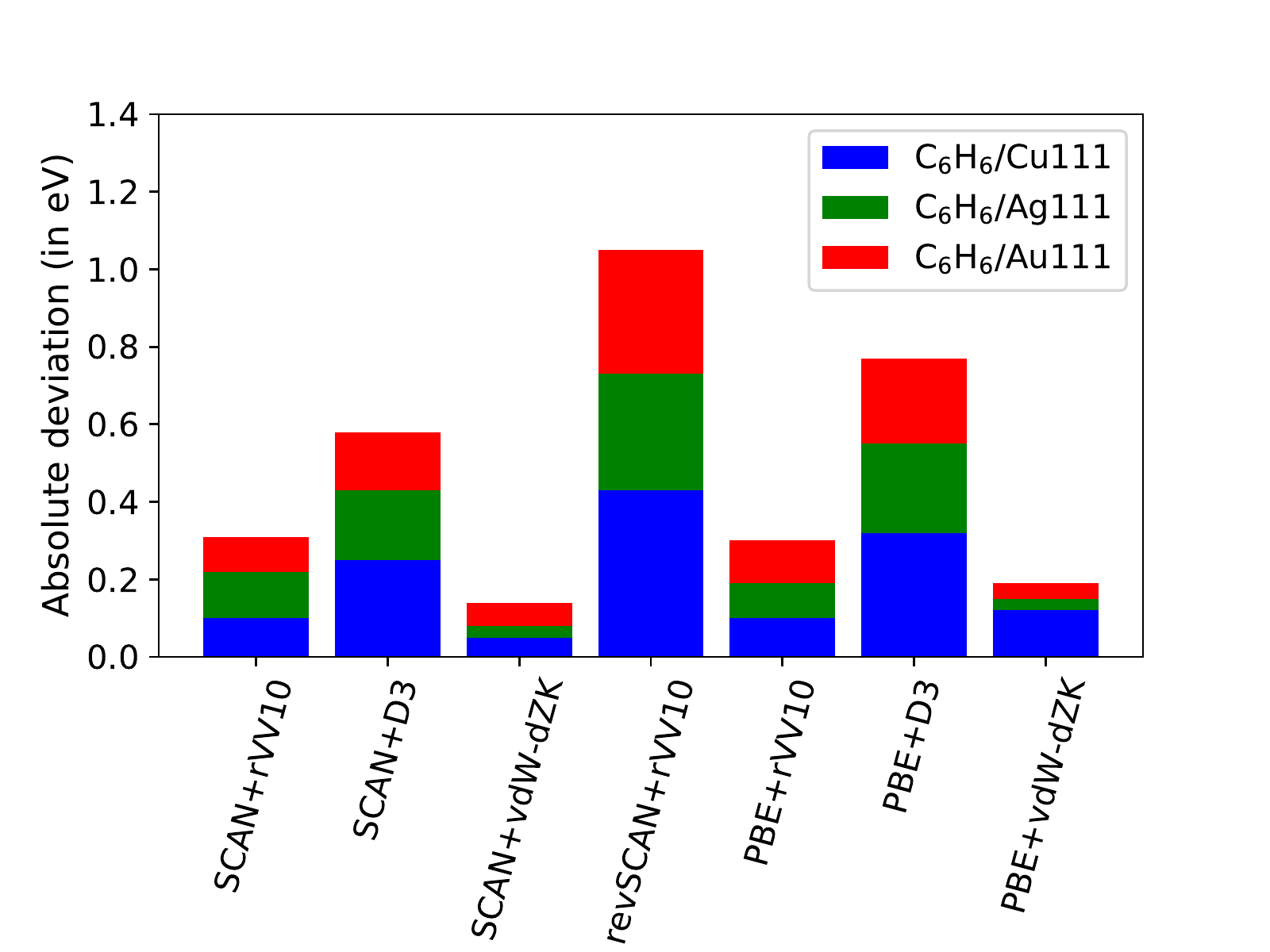}
    \caption{Figure shows the absolute deviation between adsorption energies (in eV) of benzene (C$_6$H$_6$) on the (111) surfaces of Cu, Ag, and Au calculated using various methods studied in our work and the experimental value obtained from the complete analysis.\cite{king1975thermal} We note that the uncertainities (not shown in figure) in the experimental values\cite{maass2018binding} are $\pm$0.04 eV, $\pm$0.05 eV, and $\pm$0.03 eV for C$_6$H$_6$ on the (111) surfaces of Cu, Ag, and Au respectively. }
    \label{fig:my_label}
\end{figure}
Finally, based on the molecular orbital density of states (MODOS) analysis by Jiang \textit{et al}.,\cite{jiang2016aromatic} the d-band center for benzene on Ag(111) lies farther away from the Fermi-level than for benzene on Au(111) surface. From the perspective of d-band center theory,\cite{hammer1995electronic,hammer1995gold,hammer2000theoretical} this means that Au(111) should be more reactive than Ag(111) for benzene. The experimental adsorption energies obtained from the so-called complete analysis of TPD measurements support this prediction of d-band center theory too. In this regard, SCAN+vdW-dZK and PBE+vdW-dZK both appear to be qualitatively accurate as well. Though the adsorption energies predicted from methods like PBE+MBD, HSE+MBD, and optPBE+vdW-DF were reasonable, all those methods predict almost equal adsorption energies of benzene on Ag(111) and Au(111) against the trend predicted by the experiment and the d-band center theory.\\
\begin{table}[h!]
\caption{Adsorption distance (in {\AA}) of the molecule benzene (C$_6$H$_6$) when placed on the most stable site (hcp-30) over the (111) surfaces of the coinage metals using different methods. Adsorption distance from various previous studies is also reported here for comparison.}
\begin{tabular}{lcccc}
\hline \hline
    & C$_6$H$_6$/Cu(111) & C$_6$H$_6$/Ag(111) & C$_6$H$_6$/Au(111) &  \\
\hline
PBE           & 3.40       & 3.44       & 3.42       &  \\
PBE+rVV10     & 3.01       & 3.02       & 3.01       &  \\
PBE+D3        & 2.95       & 2.97       & 2.97       &  \\
PBE+vdW-dZK   & 3.13       & 3.23       & 3.22       &  \\
PBE+vdW\cite{carrasco2014insight}          & 3.04           & 3.14           & 3.21           &  \\
PBE+vdW$^{surf}$\cite{jiang2016aromatic}          & 2.83           & 2.97           & 3.05           &  \\
B86bPBE+XDM\cite{christian2016surface}   & 2.71       &3.03        &3.15        &  \\
optPBE+vdw-DF\cite{yildirim2013trends}        & 3.14      & 3.23      & 3.21      &  \\
SCAN          & 3.01       & 3.10       & 3.18       &  \\
SCAN+rVV10    & 2.95       & 2.98       & 3.01       &  \\
SCAN+D3       & 2.90       & 3.00       & 2.99       &  \\
SCAN+vdW-dZK  & 2.85       & 3.02       & 3.06       &  \\
revSCAN+rVV10 & 2.66       & 2.93       & 2.95       &  \\
Expt\cite{liu2015quantitative}          & -           & 3.04$\pm$0.02          & -           &  \\
\hline \hline
\end{tabular}
\end{table}
$~~$Although adsorption energies over different metallic surfaces are close to each other, Reckien \textit{et al}.\cite{reckien2014theoretical} argue that the vertical adsorption distances are similar for Ag(111) and Au(111), while relatively shorter for Cu(111). This ordering of adsorption distances is consistent with the similar vdW radii of Ag and Au, both being larger than that of Cu. To the best of our knowledge, we have experimental data for adsorption distance available\cite{liu2015quantitative} only for benzene over Ag(111) based on the normal-incidence x-ray standing wave (NIXSW) measurements, as shown in Table II. Though methods like SCAN+rVV10, PBE+D3, and PBE+rVV10 predict adsorption distances for benzene over Ag(111) closer to the experimental values, they predict similar adsorption distance for different metallic surfaces. SCAN, revSCAN+rVV10, and PBE+vdW-dZK get that order right but, they are quantitatively not close enough to the experimental values. Getting the adsorption distances, orientations, and energies right are challenges even for most vdW-corrected density functionals as they require a delicate balance between short- and long-range correlations. The popular method optPBE+vdW-DF\cite{yildirim2013trends} yields the adsorption energies very close to the experimental values but predicts longer adsorption distances. However, methods like SCAN+D3\cite{dft+d3-bj} predict adsorption distances within the error bar of experimental value, yet overestimate the adsorption energies. Very few methods like B86bPBE+XDM,\cite{christian2016surface} PBE+vdW$^{surf},$\cite{jiang2016aromatic,maass2018binding} and PBE+MBD\cite{jiang2016aromatic,maass2018binding} give a reasonable description of the aforementioned challenges. The orientations and adsorption distances of benzene over different metallic surfaces predicted by some of the better-performing methods in our work are as shown in Fig. 3. PBE+vdW-dZK is not as impressive for adsorption distances as it was for adsorption energies, but SCAN+vdW-dZK stands out as the best overall performer. Apart from predicting the correct adsorption sites and orientations of the adsorbate molecule, SCAN+vdW-dZK yields the adsorption energies and distances within the narrow error bar of the available experimental values.
\begin{figure}[h!]
    \centering
    \includegraphics[scale=0.3]{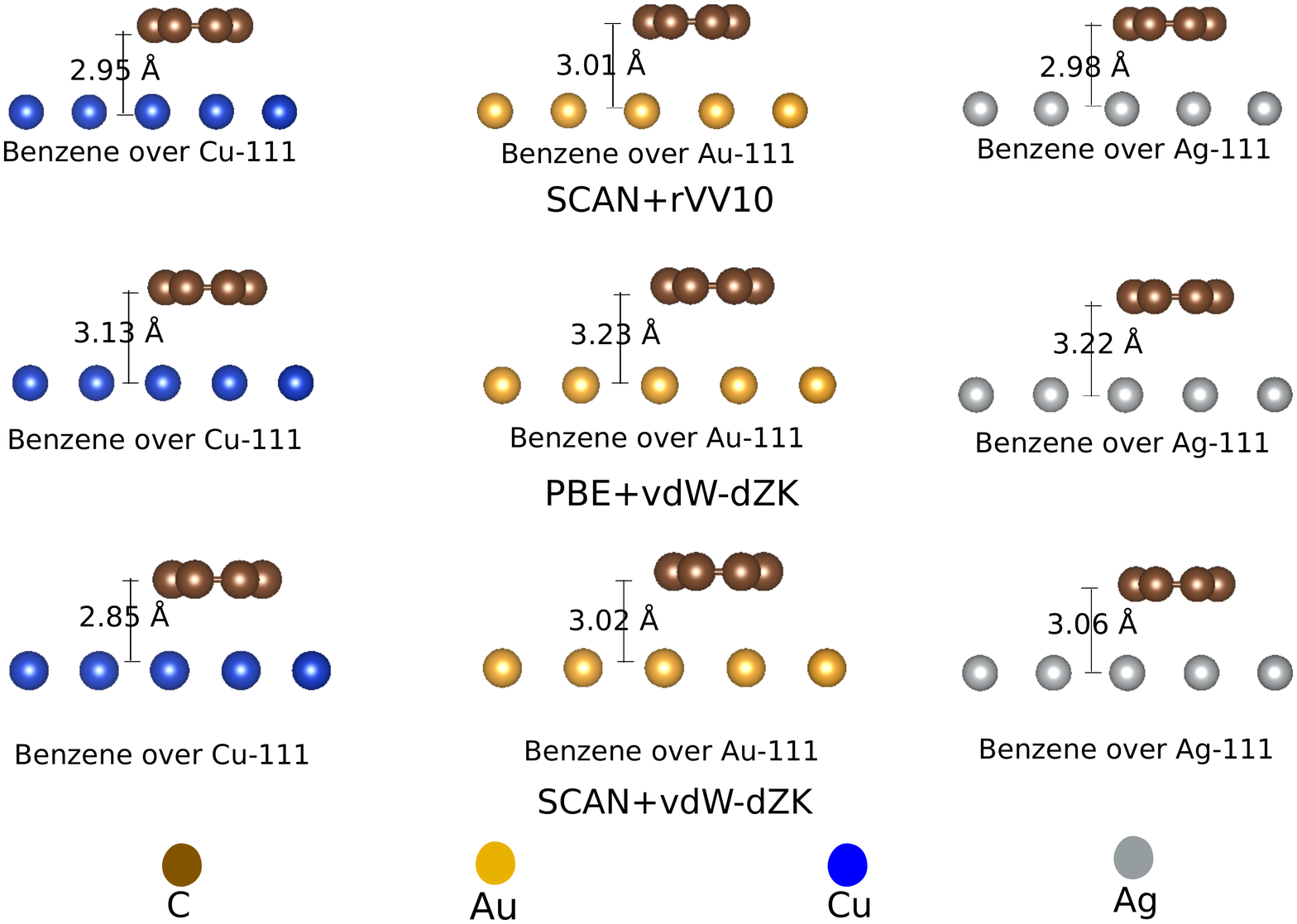}
    \caption{The figure shows the adsorption distance of benzene over the (111) surfaces of Cu, Ag, and Au when placed at the most stable site (hcp-30). For convenience, only the results of the three better performers, namely PBE+vdW-dZK, SCAN+vdW-dZK, and SCAN+rVV10, are displayed.}
    \label{fig:my_label}
\end{figure}
\subsection{Thiophene over Cu(111), Ag(111) and Au(111)}
In our previous work,\cite{adhikari2020molecule} we had demonstrated how the combination of non-local functional rVV10 combined with metaGGAs failed to deliver for adsorption of the polar molecule thiophene over the (111) surfaces of the coinage metals. We also showed how PBE+vdW-dZK was able to predict adsorption energies, orientation, and distances better than many studied and existing methods. However, in the previous section, we showed that the PBE+vdW-dZK predicted adsorption energies of benzene over coinage metals reasonably well but yielded longer adsorption distances, where SCAN+vdW-dZK was reasonable in both areas. Here we compare the performance of SCAN+vdW-dZK for thiophene over the (111) surfaces of coinage metals.\\
Experiments\cite{milligan2001complete,su2003phase,rousseau2002structure} predict the ontop position of the sulfur atom in the thiophene as the most stable site of adsorption. This site is close to the fcc-45\cite{ase-paper} site of the center of the aromatic ring of the thiophene. Based on the given coverage, SCAN+vdW-dZK, in agreement to PBE+vdW-dZK, SCAN+rVV10, revSCAN+rVV10, PBE+rVV10, PBEsol+rVV10, predicts this fcc-45 site as the most stable site of adsorption.\\
$~~~~$While experiments\cite{su2003phase,dishner1996formation,vaterlein2000orientation} demonstrate thiophene to lie flat over the (111) surfaces of Ag and Au, strong coverage dependent tilting is found\cite{milligan2001complete} over the Cu(111) surface increasing from an angle 12$^{0}\pm2^{0}$ to 25$^{0}\pm4^{0}$, when the coverage varies from 0.03 ML to 0.1 ML. Most of the theoretical studies\cite{tonigold2010adsorption,christian2016surface,adhikari2020molecule} predict thiophene to lie flat even over the Cu(111) surface. SCAN+vdW-dZK, in agreement with the experiments, predicts the flat orientation of thiophene over Ag(111) and Au(111). However, the predicted tilting angle of 7$^0$ over Cu(111) is slightly less than the tilt angles observed for 0.03 ML coverage. Similar tilting was predicted by PBE+vdW$^{surf}$.\cite{maurer2016adsorption} Though the tilting angle predicted by PBE+vdW-dZK\cite{adhikari2020molecule} is close to the values demonstrated by the experiment, Maurer \textit{et al}.,\cite{maurer2016adsorption} argue that the tilting angles predicted by experiment at finite temperature is larger due to the anharmonicity of the adsorbate-substrate bond.\\
\begin{table}[h!]
\setlength{\tabcolsep}{0.4pt}
\caption{substrate-sulfur distance (in {\AA}) and adsorption energy (in eV) of thiophene placed on the most stable site (fcc-45) over the (111) surfaces of the coinage metals using different methods.}
	\resizebox{\columnwidth}{!}{%
\begin{tabular}{lcccccc}
\hline \hline
             & \multicolumn{2}{l}{~~~C$_4$H$_4$S/Cu(111)} & \multicolumn{2}{l}{C$_4$H$_4$S/Ag(111)~~~~~} & \multicolumn{2}{l}{C$_4$H$_4$S/Au(111)} \\
             \hline
             & d(Cu-S)         & E$_{ad}$           & d(Ag-S)         & E$_{ad}$            & d(Au-S)         & E$_{ad}$           \\ \hline
PBE+rVV10\cite{adhikari2020molecule}    & 2.88            & -0.61         & 3.00               & -0.55         & 2.98            & -0.63         \\
PBE+vdW$^{surf}$\cite{maurer2016adsorption}  & 2.78            & -0.82         & 3.17            & -0.72         & 2.95            & -0.77         \\
PBE+D2\cite{tonigold2010adsorption}        & 2.40            & -0.81         &         -        &         -      & 2.75            & -1.24         \\
PBE+vdW-dZK\cite{adhikari2020molecule}  & 2.57            & -0.60          & 3.16            & -0.50          & 3.23            & -0.56         \\
B86bPBE+XDM\cite{christian2016surface}      & 2.99            & -0.62          & 3.04            & -0.67          & 2.99            & -0.66          \\
PBEsol+rVV10\cite{adhikari2020molecule}    & 2.19            & -1.22         & 2.68               & -0.93         & 2.59            & -1.06         \\
SCAN+vdW-dZK  & 2.66            & -0.69         & 3.00            & -0.62          & 2.98            & -0.69         \\
Expt         & 2.62$\pm0.03$\cite{milligan1998nixsw}            & -0.66\cite{milligan2001complete}         &      -           & -0.52\cite{vaterlein2000orientation}         &      -           & -0.68\cite{liu2002chemistry} \\
\hline \hline
\end{tabular}}
\end{table}
\begin{figure}[h!]
    \centering
    \includegraphics[scale=0.6]{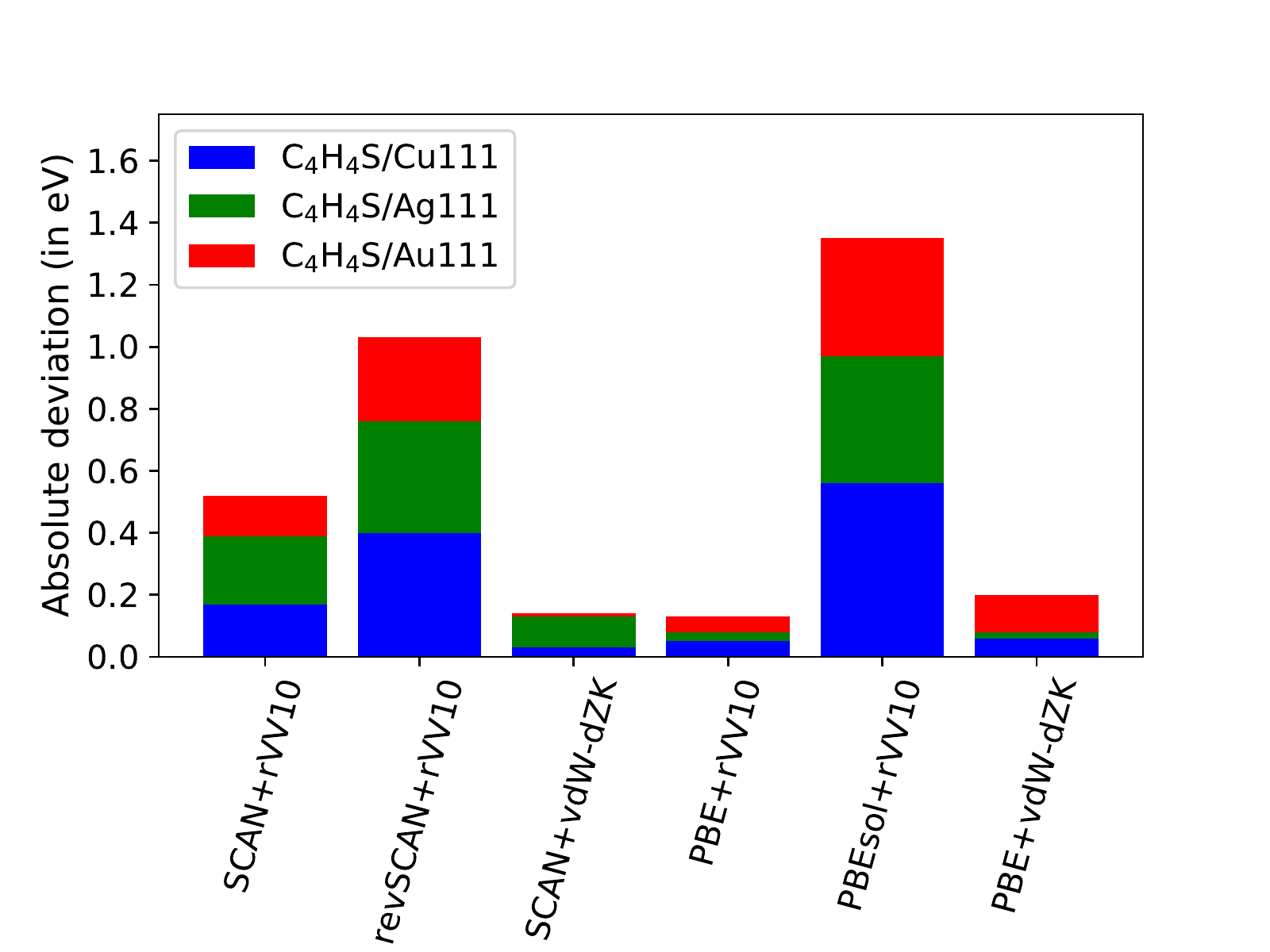}
    \caption{Figure shows the absolute deviation between adsorption energies (in eV) of thiophene (C$_4$H$_4$S) on the (111) surfaces of Cu, Ag, and Au calculated using various methods studied in our work and the experimental value\cite{milligan2001complete,liu2002chemistry,vaterlein2000orientation} obtained using the Redhead Analysis.\cite{redhead1962thermal}}
    \label{fig:my_label}
\end{figure}
$~~$Table III displays the sulfur-substrate distance (in {\AA}) and the adsorption energy (E$_{ad}$) [in eV] of thiophene over different substrate metals. The experimental adsorption energies are obtained from the Redhead analysis\cite{redhead1962thermal} of the TPD measurements \cite{milligan2001complete,liu2002chemistry,vaterlein2000orientation} using 10$^{13}$ s$^{-1}$ as the frequency parameter. We note that for a given TPD measurement, the choice of frequency parameter from 10$^{12}$ s$^{-1}$ to 10$^{15}$ s$^{-1}$ can lead to differences up to 0.2 eV. But, since thiophene is a small molecule, our choice of frequency parameter, we believe, is a reasonable one. We also note that this frequency parameter is consistent with the choice made by Christian \textit{et al}. \cite{christian2016surface} for Redhead analysis on similar work. Unfortunately, we have the substrate-sulfur distance available only for thiophene on Cu(111) based on S K-edge X-ray-absorption fine structure measurements.\cite{imanishi1998structural,milligan1998nixsw} In Fig. 4, we display the absolute deviation of the adsorption energies predicted by various methods in our work from the experimental values\cite{milligan2001complete,liu2002chemistry,vaterlein2000orientation} obtained using the Redhead analysis.\cite{redhead1962thermal} While popular methods like PBE+D2\cite{tonigold2010adsorption} overestimate E$_{ad}$ and predict shorter distances, methods like B86bPBE+XDM\cite{christian2016surface} and PBE+rVV10\cite{adhikari2020molecule} predict reasonable adsorption energies but yield longer distances. PBE+vdW-dZK predicts the substrate-sulfur distance for thiophene over Cu(111) closer to the experimental value, but it slightly underestimates the adsorption energies. SCAN+vdW-dZK, though it slightly overestimates the adsorption energy of thiophene on Ag(111), is the best overall performer (see Figures (S4-S6) in the Supplementary Materials for the details of SCAN+vdW-dZK calculations). Fig. 5 displays the orientation and sulfur-adsorbate distance of thiophene over the (111) surfaces of coinage metals for the three better-performing methods.\\
\begin{figure}[h!]
    \centering
    \includegraphics[scale=0.3]{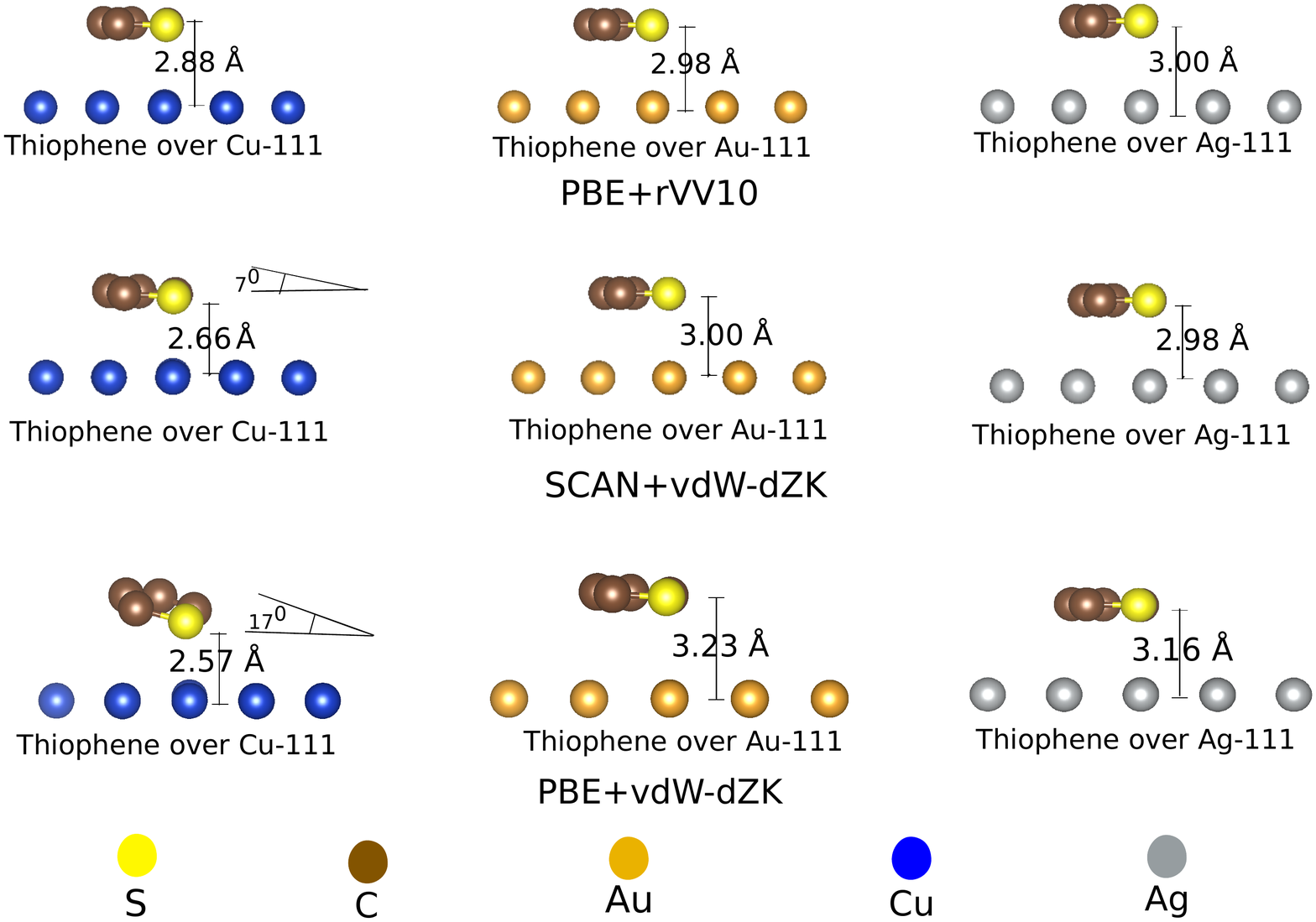}
    \caption{Figure shows the adsorption distance and tilt angle of thiophene over the (111) surfaces of Cu, Ag, and Au placed at the most stable site of adsorption (fcc-45). For convenience, results of the three better performers namely PBE+vdW-dZK, SCAN+vdW-dZK, and PBE+rVV10, are shown.}
    \label{fig:my_label}
\end{figure}

\subsection{Xenon (Xe) over Cu(111) and Ag(111)}
As a further test of vdW-dZK model, we assessed the adsorption energy and distance of Xe over the (111) surfaces of Cu and Ag. The binding of Xe with the substrate metal surfaces is a typical physisorption case dominated by vdW interaction. We compare the results from the vdW-dZK methods to the adsorption energies and distances reported by other existing vdW-corrected methods in Table IV. The experimental values tabulated here are from the so-called best estimate,\cite{vidali1991potentials} and the review work of Diehl \textit{et al}.\cite{diehl2004adsorption} (see the references therein for more details on experimental results).
\begin{table}[h!]
\setlength{\tabcolsep}{4pt}
\caption{Xe-substrate distance (in {\AA}) and adsorption energy (in meV) of Xe placed on the most stable site (ontop) over Ag(111) and Cu(111).}
	\resizebox{\columnwidth}{!}{
	\begin{tabular}{lcccc}
	\hline \hline
             & \multicolumn{2}{c}{Xe/Cu(111)} & \multicolumn{2}{c}{Xe/Ag(111)} \\
             \hline
             & E\_ad              & d(Xe-Cu)      & E\_ad              & d(Xe-Ag)      \\
             \hline
PBE          & -21                & 4.25      & -21                & 4.27      \\
PBE+vdW\cite{ruiz2016density}          & -335                & 3.48      & -244                & 3.60      \\
PBE+vdW$^{surf}$\cite{ruiz2016density}          & -248                & 3.46      & -237                & 3.57      \\
PBE+vdW-dZK  & -252               & 3.29      & -275               & 3.37      \\
PBEsol+rVV10s\cite{terentjev2018dispersion}  & -165               & 3.5      & -186               & 3.4      \\
vdW-DF\cite{silvestrelli2016van}          & -184                & 3.97      & -180                & 4.08      \\
vdW-DF2\cite{silvestrelli2016van}          & -157                & 4.01      & -154                & 4.00      \\
SCAN         & -90                & 3.61      & -93                & 3.57      \\
SCAN+vdW-dZK & -187               & 3.47      & -197               & 3.49      \\
Expt\cite{vidali1991potentials,diehl2004adsorption}         & -(183$\pm$10)      & 3.60$\pm$0.08       & -(211$\pm$15)      & 3.60$\pm$0.05\\ \hline \hline 
\end{tabular}}
\end{table}
Although all reported methods agree to the ontop site as the most stable adsorption-site, they predict significantly different adsorption energies and distances. While SCAN without any vdW correction gives the best estimate of the adsorption distances, it significantly underestimates the adsorption energies. In contrast, the non-local correlation functional vdW-DF gives a better description of adsorption energies but yields longer adsorption distances.\cite{silvestrelli2016van} Although the performance of PBE+vdW is mixed, PBE+vdW$^{surf}$ gives a reasonable estimate of both adsorption energies and distances.\cite{ruiz2016density} While PBE+vdW-dZK performs on similar lines with PBE+vdW$^{surf}$ for the adsorption energy of Xe over Cu(111), it gives a slightly overestimated value of the adsorption energy for Xe over Ag(111). PBE+vdW-dZK predicts slightly shorter adsorption distances for Xe over both the surfaces as well. Although PBEsol+rVV10s\cite{terentjev2018dispersion} performs better than the most methods discussed here, SCAN+vdW-dZK stands out overall by predicting both adsorption energies and equilibrium distances closer to the error bars of the available experimental values\cite{vidali1991potentials,diehl2004adsorption} (see Figures S7 and S8 in the Supplementary Materials for the details on PBE+vdW-dZK and SCAN+vdW-dZK calculations).  

 \subsection{Performance of the rVV10-based methods}
In our present study, we utilized a nonlocal correlation functional rVV10\cite{rvv10} paired with PBE, PBEsol, SCAN, and revSCAN to include the long-range vdW effects required for the accurate description of the organic molecule adsorbed over the metallic surfaces. We noticed a strong discrepancy in the description of the molecule-metal system between different rVV10 based methods.  While PBE+rVV10 slightly underestimates the adsorption energies for benzene, it gives a reasonable description in the case of thiophene. SCAN+rVV10 overestimates in the case of adsorption of both molecules. Some recent studies\cite{tang2020density,yu2020different} suggest SCAN+rVV10 to be overestimating in general. When combined with revSCAN,\cite{mezei2018simple} rVV10 is significantly overestimating the adsorption energies of those molecules over the metallic surfaces. Although revSCAN\cite{mezei2018simple} was designed to remove the intermediate-range interactions of SCAN, when paired with rVV10, it proved to be over-correcting. PBEsol+rVV10 yields almost double the adsorption energies of thiophene over metallic surfaces compared to PBE+rVV10.\cite{adhikari2020molecule} But, rVV10 is not the only nonlocal functional to demonstrate such strong dependence on the base functional to which it combines. A noticeable dependence on base functionals was reported for nonlocal correlation functionals such as vdW-DFs,\cite{dion2004van,vdwdf2} as well. When the revPBE\cite{revPBE} base functional proposed for vdW-DF\cite{dion2004van} was replaced by the revised version\cite{rpw86} of PW86\cite{pw86} for vdW-DF2,\cite{vdwdf2} it yielded better accuracy for molecular complexes. While both these base functionals\cite{yildirim2013trends} underestimated the adsorption energies for benzene on coinage metals, the pairing with less repulsive base functionals\cite{optPBE} like optPBE and optB88 gave a better description.\\
$~~~~$rVV10, unlike vdW-DFs, is much more flexible to pair with the base functional and does so with a set of parameters, namely `b' and `C.'  The former controls the short-range damping while the latter controls the accuracy of the C$_6$ coefficient at large separation. Most studies regarding parameterization\cite{rvv10,peng2016versatile,peng2017rehabilitation,adhikari2020molecule} of rVV10 keep the value of `C' equal to 0.0093 fixed as originally proposed by Vydrov \textit{et al}.,\cite{vydrov2010nonlocal} as changing `C' does not significantly change the binding curve.\cite{peng2016versatile} Therefore, it is usually a single parameter `b' that defines the pairing. Determination of this parameter requires fitting to well-benchmarked systems where the vdW interaction is significant, such as interaction energies of the S22 data set, the binding energy of rare-gas dimers, and interlayer binding energies of layered materials. However, we noticed a significant inconsistency in the performance of the rVV10 based methods when we determined the parameters for a particular base functional from different possible fitting methods. It is a more serious problem as this creates enough doubts over which parameterization to choose for combining rVV10 to a particular base functional for the particular problem. Fortunately for SCAN+rVV10,\cite{peng2016versatile} the `b' parameter obtained from fitting to the binding energy of rare-gas dimers produced fewer errors for other systems as well. The `b' parameter for PBE+rVV10 is 6.6 and 10.0 when fitted to the interaction energies of the S22 data set and interlayer binding energies of the layered materials, respectively.\cite{peng2017rehabilitation} However, both these values are different (b=9.7) from the parameter obtained by fitting to the argon-dimer binding energy.\cite{adhikari2020molecule} In our previous work\cite{adhikari2020molecule} we obtained b=9.7 for PBEsol+rVV10 using the argon dimer parameterization, which is significantly different from b=20 based on fitting to the interlayer binding energy of layered materials.\cite{terentjev2018dispersion} The latter choice of the parameter significantly underperformed for binding energies of the argon dimer compared to the former and was also unsatisfactory for interaction energies of the S22 data set compared to PBE+rVV10 (b=6.6) and SCAN+rVV10. \\
 \begin{table}[h!]
 \caption{Comparison of adsorption energies (in eV) of PBE+rVV10 and PBEsol+rVV10 based on separate parameterizations using the S22 data set, Ar$_2$ binding energies and interlayer binding energies of layered materials (LM) for thiophene adsorbed on the (111) surface of coinage metals.}
\begin{tabular}{lccccc}
\hline \hline
       & ref   & \multicolumn{2}{l}{PBE+rVV10} & \multicolumn{2}{l}{PBEsol+rVV10} \\
       \hline
       &       & Ar$_2$      & S22     & Ar$_2$       & LM      \\
       \hline
Cu & -0.66\cite{milligan2001complete} & -0.61         & -0.95         & -1.22          & -0.75           \\
Au   & -0.68\cite{liu2002chemistry} & -0.63         & -0.94         & -1.06          & -0.69           \\
Ag & -0.52\cite{vaterlein2000orientation} & -0.55         & -0.82         & -0.93          & -0.59          \\
\hline \hline
\end{tabular}
\end{table}
$~~~~$In Table V, we compare the adsorption energy of thiophene over the coinage metals, at the most stable site (fcc-45), from the different parameterization of rVV10 for PBE or PBEsol. We observed significant differences in the adsorption energies when different parameterizations combined rVV10 to the same base functional, illustrating the inconsistency in the performance of rVV10-based methods. The need for an improved two-parameter damping function of rVV10 is being explored by Tang, Chowdhury, and Perdew\cite{revisedrvv10}, and it would hopefully solve larger parts of this problem.
\section{Conclusion} 
In this work, we investigated the geometry and energetics of the adsorption of xenon and two widely studied prototype molecules, namely, benzene, and thiophene, over the (111) surface of the coinage metals using density functional theory (DFT).\\ 
$~~~~$Recently, the so-called complete analysis method\cite{king1975thermal} was used to analyze the temperature-programmed desorption (TPD) measurement of benzene over coinage metal surfaces\cite{maass2018binding} yielding adsorption energies within an error bar of chemical accuracy (0.04 eV). Based on our results, the recently introduced model based on Zaremba-Kohn's second-order perturbation theory (DFT+vdW-dZK) predicts the adsorption energies of benzene over metal surfaces better than any other methods reported here. In particular, SCAN+vdW-dZK stands out. Apart from yielding the correct orientation and site, this method predicts the energies and distances within the error bar of the experiment. SCAN+vdW-dZK predicts Au(111) to be more reactive than Ag(111), supporting the popular d-band center theory. SCAN+vdW-dZK is the best overall performer for xenon and thiophene over the metallic surfaces, as well. We note that due to the unavailability of the results from the complete analysis, Readhead's analysis, \cite{redhead1962thermal} which is also widely used but not as accurate as the former, was utilized here to analyze the TPD measurements of thiophene adsorbed over the metallic surfaces. \\ 
$~~~~$Finally, through the example of thiophene adsorbed on metal surfaces, we demonstrate that the rVV10 based methods are inconsistent. They predict significantly different adsorption energies of a molecule-metal system when the same base functional pairs with rVV10 through different possible parametrizations. The current rVV10 is essentially a pairwise form in which the screening effects of metal substrates are not included explicitly in the molecule-metallic surface interactions. The vdW-dZK model incorporates such screening effects explicitly via the C$_3$ and C$_5$ nonlocal terms making the model more accurate to describe the molecule-metallic surface interactions. Although the vdW-dZK model combines with PBE or SCAN through a single parameter 'b', this model relies on the accurate static dipole polarizabilities as an input parameter. In our present work, we show that SCAN+vdW-dZK is slightly better than PBE+vdW-dZK on the overall performance. The adsorption distances predicted by SCAN+vdW-dZK show better agreement with the experimental values compared to PBE+vdW-dZK. However, there are no significant differences in the adsorption energies predicted by these methods. Despite the vdW-dZK methods combining with two different functionals PBE and SCAN, it is remarkable that the adsorption energies both these methods predict are close to each other, and importantly, both close to the experimental values.\\ 

\begin{acknowledgements}
S.A. and A.R. acknowledge the support of the U.S. Department of Energy, Office of Science, Office of Basic Energy Sciences, as part of the Computational Chemical Sciences Program under Award No. DE-SC0018331. H.T. acknowledges support from the DOE
Office of Science, Basic Energy Sciences (BES), the U.S. Department of Energy, under Grant No. DE-SC0018194. The work of N.K.N. was supported by the National Science Foundation (NSF), under the grant number DMR-1553022. This research includes calculations carried out on HPC resources supported in part by the NSF through major research instrumentation grant number 1625061 and by the U.S. Army Research Laboratory under contract number W911NF-16-2-0189. We thank Prof. John P. Perdew for his valuable help throughout.\\~\\~\\~\\
\end{acknowledgements}

%

\end{document}